# Deep convolutional neural networks for lung nodule detection: improvement in small nodule identification

Sunyi Zheng*, Ludo J. Cornelissen, Xiaonan Cui, Xueping Jing, Raymond N. J. Veldhuis, Senior Member, IEEE, Matthijs Oudkerk, and Peter M.A. van Ooijen, Senior Member, IEEE

*Abstract*— *Objective:* In clinical practice, small lung nodules can be easily overlooked by radiologists. The paper aims to provide an efficient and accurate detection system for small lung nodules while keeping good performance for large nodules. *Methods:* We propose a multi-planar detection system using convolutional neural networks. The 2-D convolutional neural network model, U-net++, was trained by axial, coronal, and sagittal slices for the candidate detection task. All possible nodule candidates from the three different planes are combined. For false positive reduction, we apply 3-D multi-scale dense convolutional neural networks to efficiently remove false positive candidates. We use the public LIDC-IDRI dataset which includes 888 CT scans with 1186 nodules annotated by four radiologists. *Results:* After ten-fold cross-validation, our proposed system achieves a sensitivity of 94.2% with 1.0 false positive/scan and a sensitivity of 96.0% with 2.0 false positives/scan. Although it is difficult to detect small nodules (i.e. < 6 mm), our designed CAD system reaches a sensitivity of 93.4% (95.0%) of these small nodules at an overall false positive rate of 1.0 (2.0) false positives/scan. At the nodule candidate detection stage, results show that a multi-planar method is capable to detect more nodules compared to using a single plane. *Conclusion:* Our approach achieves good performance not only for small nodules, but also for large lesions on this dataset. This demonstrates the effectiveness and efficiency of our developed CAD system for lung nodule detection. *Significance:* The proposed system could provide support for radiologists on early detection of lung cancer.

(Corresponding author: Sunyi Zheng.)
S. Zheng, L.J. Cornelissen, X. Jing and P.M.A. van Ooijen are with the University of Groningen, University Medical Center Groningen, Department of Radiation Oncology, 9713 GZ Groningen, The Netherlands (e-mail: s.zheng@umcg.nl; l.j.cornelissen@umcg.nl; x.jing@umcg.nl; p.m.a.van.ooijen@umcg.nl)
X. Cui is with the University of Groningen, Faculty of Medical Science University Medical Center Groningen, Department of Radiology, 9713 GZ Groningen, The Netherlands and also with the Tianjin Medical University Cancer Institute and Hospital, National Clinical Research Centre of Cancer, Department of Radiology, 300060 Tianjin, China (e-mail: x.cui@umcg.nl)
R. N. J. Veldhuis is with the University of Twente, Faculty of Electrical Engineering, 7500 AE Enschede, The Netherlands (e-mail: r.n.j.veldhuis@utwente.nl)
M. Oudkerk is with the University of Groningen, Faculty of Medical Science, 9713 AV Groningen, The Netherlands (e-mail: m.oudkerk@umcg.nl)

## I. INTRODUCTION

Lung cancer is one of the most malignant cancers, and is a leading cause of death among both men and women [1, 2]. It has been predicted that around 25% of all cancer deaths in the U.S. in 2019 are due to lung cancer [3]. Early detection of lung cancer can give better treatment alternatives to patients and increase their survival chances [4]. To improve early diagnosis, lung cancer screening trials, such as the National Lung Screening Trial (NLST) [5] and the Dutch-Belgian Randomized Lung Cancer Screening Trial (NELSON) [6], have been implemented.

Although the implementation of lung cancer screening reduces the mortality rate of patients, it results in a heavy workload for radiologists. Computer-aided detection (CAD) systems could play an essential role in assisting radiologists to find nodules efficiently. A CAD system generally consists of two stages: Suspicious candidate detection and false positive reduction. The aim of any CAD system for lung nodule detection is to reach a high sensitivity with a low false positive (FP) rate. However, CAD systems still have not been widely used in clinical practice for various reasons, including lack of reimbursement and low sensitivity or high false positive rates of the available systems [7, 8]. The challenges of this task are mainly the large variety in nodule morphology and the detection of small nodules, which are easily missed.

With the development of artificial intelligence algorithms and the abundance of computational power, a large number of deep learning techniques have been successfully used in image processing fields. For example, Ronneberger et al. [9] proposed the U-net algorithm for biomedical image segmentation, which showed good performance in the IEEE International Symposium on Biomedical Imaging (ISBI) cell tracking challenge. The U-net algorithm is widely used for segmentation tasks throughout the literature ever since [10-13]. Variations on this architecture were soon proposed, such as the improved model U-net++ from Zhou et al. [14], which modifies the skip connections between encoder and decoder pathways in the network. This should reduce the semantic gap between feature maps from the decoder and encoder paths, which makes training more efficient. Considering network architectures for image classification, Tan et al. [15] demonstrated that by scaling depth, width, and resolution, Efficient-Net becomes more accurate for object classification assessed on the CIFAR-10 dataset. Inspired by dense convolution networks [16], Huang et al. [17] developed a more effective architecture for image classification by adding multi-scale blocks.

Meanwhile, various authors have reported automatic lung nodule detection algorithms using deep learning. In the effort to minimize false negatives and false positives, Wang et al. [18] proposed a nodule-size-adaptive model that can measure the nodule sizes, types and locations from 3-D images. Moreover, Dou et al. [19] used 3D convolutional neural networks to extract multilevel contextual information to reduce false positives, while Xie et al. [20] utilized 2D convolutional neural networks for false positive reduction. Another approach by Setio et al. [21] combined the predictions from seven independent nodule detection systems and five false positive reduction systems. Some of the detection systems were developed for specific types of nodules. In addition, Ozdemir et al. [22] developed an end-to-end system for nodule detection by utilizing 3D convolutional neural networks based on V-net. Furthermore, Zhang et al. [23] applied constrained multi-scale Laplacian of Gaussian filters to localize potential nodule candidates and a densely dilated 3D convolutional neural network to reduce false positives.

In our previous work, we followed one of the clinical procedures: Maximum intensity projection. With projected images as input, convolutional neural networks (CNNs) were employed to identify nodule candidates [24]. Nodule cubes with various sizes were extracted for reduction of false positives. The results showed that using maximum intensity projection can improve the performance of deep learning-based CAD for lung nodule detection. In this work, we again attempted to learn from the clinical procedures, and tried to identify those aspects that could be mimicked in algorithm design. In particular, for clinical evaluation of a scan, radiologists would commonly take the axial, coronal and sagittal planes into account, rather than solely the axial plane. However, previous work on nodule detection is mostly based on the axial plane alone [18, 21, 23, 24]. The performance and influence on both the axial, coronal and sagittal plane for nodule detection in a deep learning-based CAD system has not been explored. More importantly, radiologists' sensitivity on small nodules is not high on CT scans in clinical practice [25-27].

The key contributions of this paper are as follows. (1) Although it is difficult to detect small nodules (i.e. nodules with a diameter < 6 mm), our designed CAD system achieved good performance on these small nodules. (2) Considering the axial plane, the coronal plane and the sagittal plane, we developed an automatic nodule identification system based on multi-planar convolutional neural networks using transfer learning. (3) We also explored the performance and influence of each plane for nodule detection in a CAD system. Combined results from three planes on the detection performance were reported. To further boost the performance, results of the proposed system on ten mm axial maximum intensity projection-based slices were merged since the ten mm slices had the highest detection rate and a relatively low false positive rate found in the previous work [28]. (4) Based on convolutional neural networks, a multi-scale dense architecture was applied to exclude suspicious candidates. Features at low or high levels can be efficiently extracted and concatenated for prediction. (5) In the false positive reduction stage, we evaluated the effect of two factors: Segmentation of lung parenchyma and the region of interests of input data.

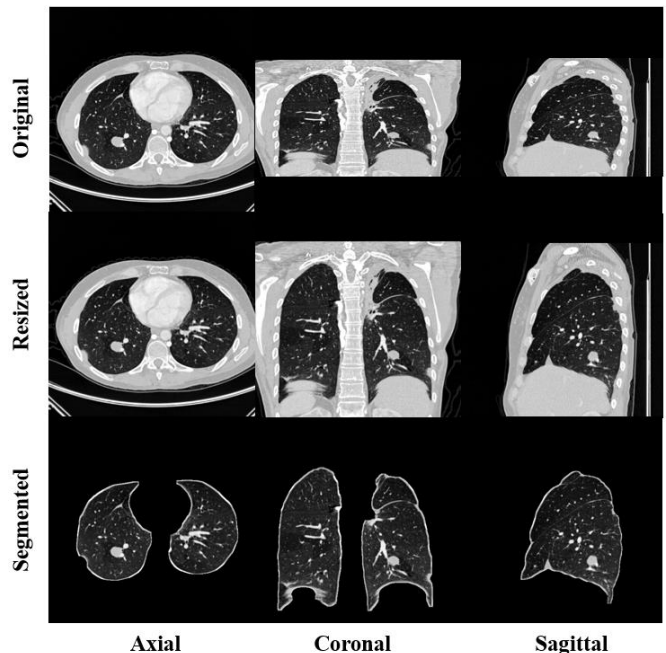

Fig. 1. Examples of preprocessing for axial, coronal, and sagittal slices. The first column is original CT data and the second column represents slices after resizing by interpolation. In the last column, segmented lung parenchyma in various directions is applied as input for training convolutional neural networks later.

## II. MATERIALS

The public dataset named Lung Image Database Consortium and Image Database Resource Initiative (LIDC/IDRI) [29] was established by seven academic centers and eight medical imaging companies. The database has 1018 CT scans and the range of the slice thickness is from 0.6 mm to 5.0 mm. These scans were reviewed by four radiologists in two reading phases. In the first round, radiologists independently detected suspicious lesions and categorized them into three groups (nodules ≥ 3 mm, nodules < 3 mm, and non-nodules). Then, findings of each scan from four radiologists were collected together and individual radiologist checked every annotation again in an unblinded way.

In clinical practice, scans with low slice thickness are recommended for pulmonary nodule detection [30]. Hence, we excluded scans with slice thickness above 2.5 mm. After discarding scans without consistent slice spacing, there were 888 scans included in our study. Nodules larger than 3 mm were considered as relevant lesions according to NLST screening protocols [5]. Consequently, we selected the 1186 nodules which were accepted by at least three radiologists. Nodules ≥ 3 mm identified by the minority of radiologists, nodules < 3 mm, and non-nodules were referred as unrelated

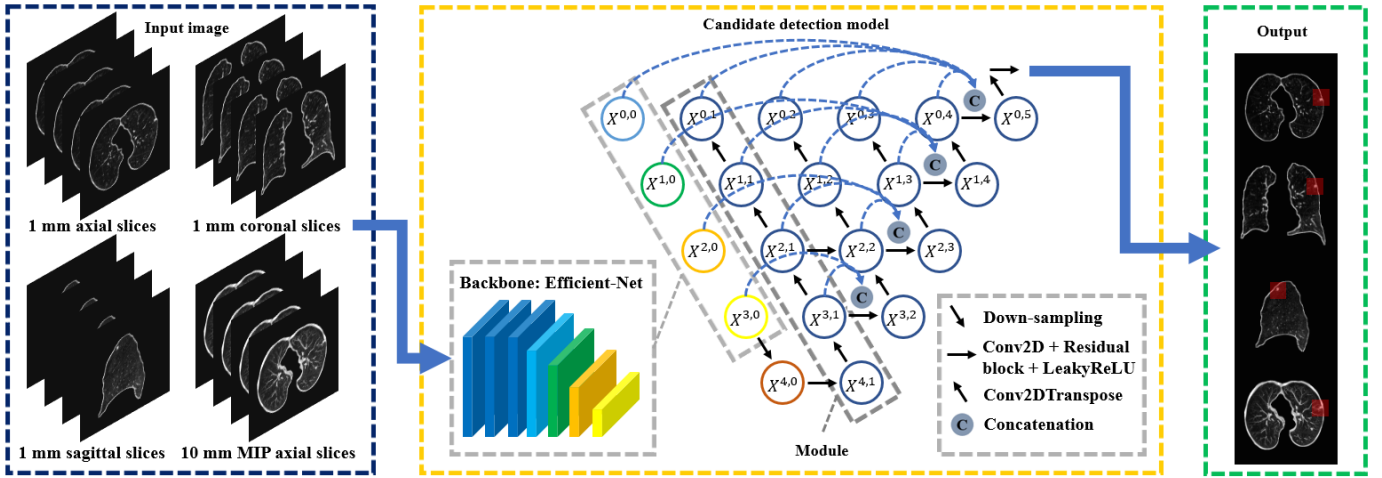

Fig. 2. The overview of our proposed candidate detection method. The 1 mm slices on axial, coronal, sagittal plane, and 10 mm axial maximum intensity projection (MIP) slices were used as input. The backbone of the detection model was based on efficient-net pre-trained on ImageNet. The proposed model efficiently extracted features not only in small receptive fields, but also large receptive fields. After prediction, suspicious findings on each plane were localized by bounding boxes.

findings.

## III. METHODS

The designed method contains two stages, namely, multi-planar nodule candidate detection and false positive reduction. We use a convolutional neural network model, U-net++, to detect potential nodule candidates on axial, coronal and sagittal planes. The backbone of the U-net++ is the Efficient-Net classification model, pre-trained on ImageNet, which efficiently extracts various basic features. The predictions from the three planes were merged to acquire a higher sensitivity. For false positive exclusion, we apply multi-scale dense convolutional neural networks to efficiently remove false positive candidates. The following sections provide more details of architectures, training progress and evaluation methodology.

### A. Data preparation

In order to convert DICOM files to PNG format, window setting was set from -1000 HU to 400 HU and pixel values range from 0 to 255. Scans in the LIDC-IDRI database have various spacing in different planes, which results in misshapen images for nodule detection. Original examples are shown in the first row of Fig. 1. To unify the data, we adopted 1 mm as the spacing value to resize the images by interpolation since thin-thickness slices improve diagnosis [31]. Moreover, segmentation of lung parenchyma can increase efficiency of training convolutional neural networks for lung nodule detection [32]. The average of pixel values in the whole slice was applied as a threshold to roughly separate lung parenchyma from the body. We removed the irrelevant information in the border and employed morphological closing to fill holes. To keep more boundary texture for detection of wall-attached nodule, morphological operation, dilation, was used. The segmentation procedure is similar to what was used in our previous work, and is described and illustrated in more detail in [24]. Segmented lung parenchyma in three planes are illustrated in the third row of Fig. 1.

### B. Multi-planar detection via transfer learning

Nodule candidate detection is a fundamental step as it is highly related to the final sensitivity of the CAD system. To achieve as high sensitivity as possible, we applied not only axial slices, but also coronal and sagittal slices for nodule candidate localization. The reason of utilizing three planes is that one nodule might be not obviously showing in one plane. To further improve detection accuracy, we combined our previous work and used 10 mm axial maximum intensity projection (MIP) based slices generated for nodule detection. The output of the candidate detection stage is constructed by using a union join from the output of four CNNs streams. More specifically, detected candidates on coronal and sagittal plane were first transformed back to axial coordinates. Potential candidates are merged if the largest radius of the candidates is smaller than 0.88 times central distance between the two. This value was determined experimentally.

With its a series of dense skip pathways between decoder and encoder networks, U-net++ shows good performance in segmentation tasks [14]. Based on U-net++, we proposed our object detection model, as shown in Fig. 2. The input slices have the size of $512 \times 512$. The architecture has two parts, namely encoder and decoder. For the encoder part, we adopt Efficient-Net [15] as backbone because it is more efficient on simple feature extraction and had the promising results on the CIFAR-100 image classification task. The model EfficientNetB4 was pre-trained on the ImageNet dataset, and its pre-trained weights were downloaded from the python package website (https://pypi.org/project/keras-efficientnets/). Using a pre-trained model based on a large dataset such as ImageNet, and then re-training (also called fine-tuning) that model on a different dataset for a different task is known as transfer learning. Transfer learning has shown good results on different tasks in the past [33-35], and the main benefits of it include that the model will already have rich feature maps prior to fine-tuning which can speed up the training and give better performance on other datasets. Efficient-Net has a compound scaling method, which results in 8 versions of the Efficient-Net. The method applies a compound coefficient μ to constrain width (w), depth (d), and resolution (r) in networks:

$$d = \alpha^\mu$$
$$w = \beta^\mu \quad (1)$$
$$r = \gamma^\mu$$
$$s.t. \; \alpha \cdot \beta^2 \cdot \gamma^2 \approx 2, \alpha \geq 1, \beta \geq 1, \gamma \geq 1$$

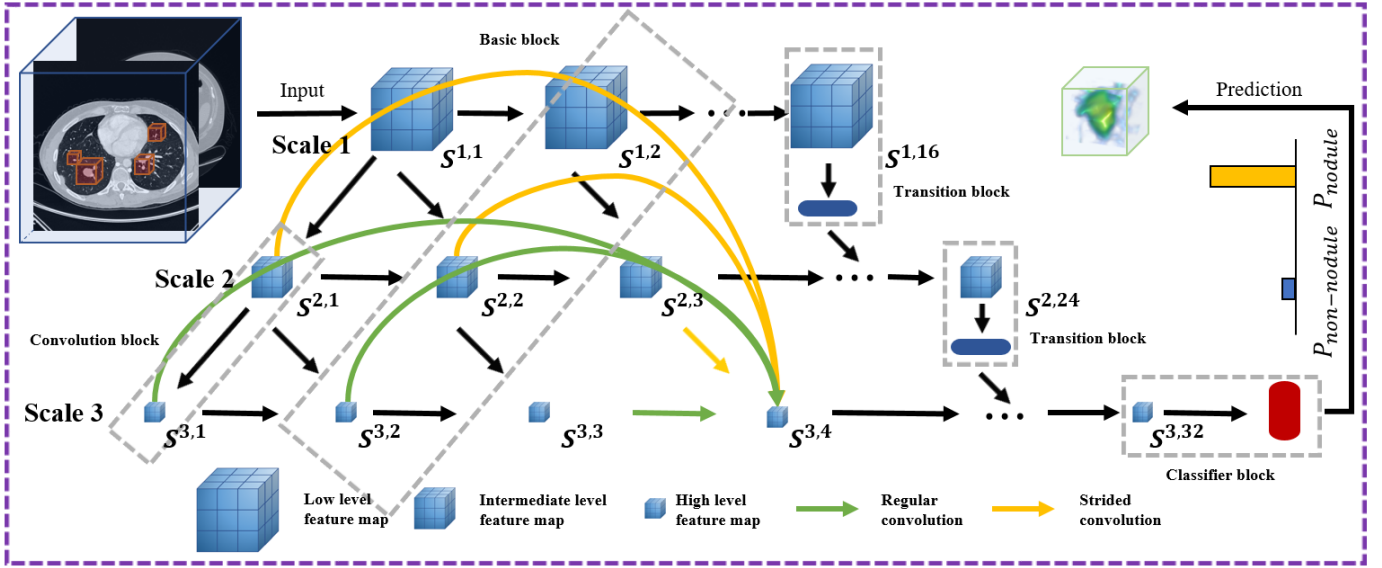

Fig. 3. The scheme for false positives reduction based on 3D multi-scale dense convolutional neural networks. The cubes are extracted from 3D volume as input. Feature maps are in three scales and the scale 1, 2, 3 has a depth of 16, 24, and 32, respectively. An example of concatenated feature maps from different levels though regular and strided convolutions is shown in scale 3 at layer 4. The classifier is in the end of scale 3, giving a probability of being a nodule for each cube. Connection between different scales and horizontal layers are not drawn explicitly.

Where α, β, and γ are constants. To avoid amount of computation more than $2^\mu$, the product of $\alpha \cdot \beta^2 \cdot \gamma^2$ is close to two. The width, depth, resolution, and the dropout rate of EfficientNetB4 that we use are 1.4, 1.8, 380 and 0.4 respectively, established by a grid search experiment [15]. The output of the backbone is connected to Leaky Rectified linear units (LeakyReLU) which is applied to prevent vanishing gradients in parts of the network [36]. The negative slope coefficient of the LeakyReLU is 0.1. Then, it is followed by a max-pooling layer and a dropout layer. The dropout rate in this architecture is 0.1. In the middle of the U-Net++, the first convolutional layer consists of 256 kernels of size 3 by 3 between $X^{4,0}$ and $X^{4,1}$. In order to increase the depth of the model, we apply two residual blocks which have 256 channels with LeakyReLU as the activation function. The decoding pathway consists of five similar modules. The first module ($X^{4,1} \to X^{3,1} \to X^{2,1} \to X^{1,1} \to X^{0,1}$) is made of four transposed 2D-convolutional layers, one concatenation layer, one dropout layer, one convolutional layer and one residual block with LeakyReLU as the activation function. More specifically, in order to revert the spatial compression, we employ four transpose 2D-convolutional layers with 128 kernels of size 3 by 3 for up-sampling [37]. Then, the concatenation layer combines related feature maps from transposed convolutional layers at previous one level and the corresponding layer in the encoding pathway (backbone: Efficient-NetB4). At each horizontal level, all concatenated feature maps are merged on the ultimate node on that level (nodes $X^{3,2}$, $X^{2,3}$, $X^{1,4}$, $X^{0,5}$). After the concatenation layer followed by one drop-out layer (drop rate: 0.1) and one convolutional layer with 128 kernels, there is a 128-channel residual block activated by LeakyReLU. For the following four modules, the number of transposed convolutional layers is reduced by one and the number of channels/kernels is halved for each subsequent module. For example, the second module is comprised by the pathway $X^{3,2} \to X^{2,2} \to X^{1,2} \to X^{0,2}$. The last module is almost the same as the fourth module. However, it does not have the concatenation layer and has one more dropout layer with the rate 0.05. The last layer is a convolutional layer with a kernel size of $1 \times 1$ and a sigmoid activation function. After prediction, suspicious findings on each plane are localized by bounding boxes.

In the training stage, each input image has at least one nodule. We augment these images through 90°, 180° and 270° rotation, horizontal-vertical flipping, and an affine transformation. The data was separated by the LUNA16 challenge into 10 folders [21]. Hence, we validate the model by 10-fold cross-validation. Every time, we leave one-fold of the data for testing. The remaining nine folds of the data are split in a training (70%) and validation (30%) set. We use a batch size of 8 and the Adam optimizer [38]. To calculate the overlap between ground truth and prediction, we apply dice as the loss function [39]. The initial learning rate is $10^{-3}$ and the minimum value is $10^{-4}$. The decreasing factor of the learning rate scheduler is set to $10^{-1}$. If the minimum validation loss does not change for 5 epochs, the learning rate decreases. The training ends when the model minimum loss on the validation set does not change for 10 epochs.

TABLE I
PERFORMANCE OF THE CAD PROGRAM USING 1 MM SLICES IN THREE DIRECTIONS AND 10 MM AXIAL MAXIMUM INTENSITY PROJECTION (MIP) IMAGES AS INPUT, AS WELL AS FUSED RESULTS AT THE NODULE CANDIDATE DETECTION. TOTAL NUMBER OF NODULES IS 1186 WITHIN 888 SCANS

| Input data | Number of detected nodules | Sensitivity (%) | False positives per scan |
|---|---|---|---|
| 1 mm axial slices | 1081 | 91.1% | 38 |
| 1 mm coronal slices | 979 | 82.5% | 33 |
| 1 mm sagittal slices | 970 | 81.8% | 40 |
| 10 mm MIP images | 1105 | 93.2% | 22 |
| Fusion 1 mm slices | 1140 | 96.1% | 98 |
| Fusion all | 1163 | 98.1% | 108 |

TABLE II
PERFORMANCE ON DETECTED NODULES IN SIZES AND TYPES AT THE NODULE CANDIDATE DETECTION STAGE. TOTAL NUMBER OF NODULES IS 1186. REGARDING TO SIZES, THERE ARE 502 NODULES (3-6 MM), 276 NODULES (6-8), 281 NODULES (8-15), 127 NODULES (≥15). IF NODULES ARE CATEGORIZED BY TYPES ACCORDING THE MAJORITY VOTES FROM FOUR RADIOLOGISTS, THERE ARE 64 GROUND-GLASS NODULES, 189 SUB-SOLID NODULES, 933 SOLID NODULES

| Nodule diameter | Nodule type | | | Total |
|---|---|---|---|---|
| | Ground-glass | Sub-solid | Solid | |
| 3-6 mm | 25 (89%) | 75 (100%) | 387 (97%) | **487 (97%)** |
| 6-8 mm | 13 (93%) | 41 (100%) | 220 (100%) | 274 (99%) |
| 8-15 mm | 18 (95%) | 48 (100%) | 211 (99%) | 277 (99%) |
| ≥15 mm | 2 (67%) | 25 (100%) | 98 (99%) | 125 (98%) |
| Total | 58 (91%) | 189 (100%) | 916 (98%) | 1163 (98%) |

TABLE III
PERFORMANCE OF THE CAD SCHEME WITH VARIED CONFIGURATIONS AT THE FALSE POSITIVE REDUCTION STAGE

| Segmentation | Region of interest | CPM |
|---|---|---|
| Segmented | Original | 0.933 |
| Unsegmented | Original | 0.937 |
| Unsegmented | Four pixels larger | 0.940 |
| Unsegmented | Eight pixels larger | 0.940 |

C. Multi-scale dense training for false positive reduction

Reduction of false positives is also essential for radiologists in clinical practice. The aim of this stage is to lower the number of nodule candidates so that fewer nodule candidates have to be manually inspected, ultimately reducing the workload of radiologists. The scheme that we propose here is based on 3D multi-scale dense convolutional neural networks [17], as shown in Fig. 3. Overall, the model has feature maps at three different scales and a maximum depth of 32 in the vertical and the horizontal direction, respectively. Green arrows indicate regular convolution operations in the horizontal path, while orange arrows represent strided convolution operations in the vertical path. Feature maps are efficiently extracted and concatenated from the results of regular convolutions on the same scale and the result of strided convolutions on the previous scale. Connection between different scales and horizontal layers are not drawn explicitly. But an example on scale of 3 at a depth of 4 with green and orange arrows is shown in Fig. 3. The network consists of thirty-two basic blocks, five transition blocks and a classifier block. The architecture starts with three convolution blocks to extract initial feature maps in three scales. Each convolution block contains a convolutional layer with a kernel size of 3×3×3 followed by batch normalization with the activation function ReLU [40]. On three scales, their numbers of filters are 32, 64, 128, and growth rates are 8, 16, 32, separately. A basic block includes three concatenation layers and five bottleneck operations that are used to reduce the number of features and improve calculation efficiency. Every bottleneck operation consists of two convolution blocks. After the bottleneck block, the number of filters is reduced by 75%. On scales 2 and 3, coarse and fine features are aggregated by concatenation from the previous and current scales. When extracting features by the strided convolution from the previous scale, the stride depth is two rather than one for a larger receptive filed. To further improve model compactness, transition blocks are designed to reduce the number of feature-maps. A transition block has a convolution block with a stride of one and a kernel size of $1 \times 1 \times 1$. The transition blocks are connected to the basic block and located at a depth of 16 and 24 in three scales. The final block is a classifier block which gives, for each input cube, the probability of containing a nodule. It has two convolutional blocks, an average pooling layer with stride of two, a flatten layer, two dense layers (128, 32 filters) and two dropout layers with the rates of 0.5 and 0.2 separately. The initializer in the convolutional layer is the he_normal [41].

Before the training session, each candidate needs to be rescaled to $32 \times 32 \times 32$. The rough size of every candidate is first estimated in the candidate detection stage, which gives a bounding box for candidates according to their diameters. However, the surrounding textural information also influences the differentiation between nodules and non-nodules for convolutional neural networks. Therefore, we experiment with two parameters that govern the availability of textural context to the false positive reduction model: 1) Whether or not the lung parenchyma is segmented and 2) size of the region of interests of the input data.

Ten-fold cross-validation is employed to evaluate the performance of the model. The same procedure as for the nodule candidate detection stage is used, with candidates from one-fold used for testing, and the nodule candidates from the remaining nine folds split for training and validation in a 70/30 ratio. The loss function is binary cross-entropy and the optimizer is Adam. The learning rate is $10^{-4}$. If the validation loss does not change for 6 epochs, the training stops.

D. Performance evaluation

At the nodule candidate detection stage, sizes and types of detected nodules from our proposed CAD system are analyzed. The nodule size is provided by the dataset and the nodules are categorized into three types according to their texture scores. In the LIDC/IDRI database, radiologists give 5 scores (1 = ground-glass, 2-4 = part-solid, 5 = solid) for nodule types. If the majority of votes is 1 and 5, the nodule type is ground-glass and solid, respectively. Otherwise, the nodule type is part-solid.

In our case, the number of true positives is much smaller than the number of false positive findings. Using the area under the ROC curve as an evaluation metric therefore does not reflect the performance of the CAD system objectively [42]. Thus, we

TABLE IV
PERFORMANCE OF OTHER COMPUTER-AIDED DETECTION SYSTEMS EVALUATED ON THE LIDC-IDRI DATABASE

| CAD system | Year | False positives per scan | | | | | | | CPM |
|---|---|---|---|---|---|---|---|---|---|
| | | 0.125 | 0.25 | 0.5 | 1 | 2 | 4 | 8 | |
| Our current work | 2020 | **0.893** | 0.917 | 0.930 | 0.942 | 0.960 | 0.966 | 0.973 | 0.940 |
| Setio et al. [21] | 2017 | 0.859 | **0.937** | **0.958** | **0.969** | **0.976** | **0.982** | **0.982** | 0.952 |
| Zhang et al. [23] | 2018 | 0.890 | 0.931 | 0.944 | 0.949 | 0.965 | 0.972 | 0.976 | 0.947 |
| Zheng et al. [24] | 2019 | 0.876 | 0.899 | 0.912 | 0.927 | 0.942 | 0.948 | 0.953 | 0.922 |
| Ozdemir et al. [22] | 2019 | 0.832 | 0.879 | 0.920 | 0.942 | 0.951 | 0.959 | 0.964 | 0.921 |
| Wang et al. [18] | 2019 | 0.788 | 0.847 | 0.895 | 0.934 | 0.952 | 0.959 | 0.963 | 0.905 |
| Dou et al. [19] | 2017 | 0.677 | 0.737 | 0.815 | 0.848 | 0.879 | 0.907 | 0.922 | 0.826 |
| Xie et al. [20] | 2019 | 0.734 | 0.744 | 0.763 | 0.796 | 0.824 | 0.832 | 0.834 | 0.790 |

The highest score of each column is shown in bold.

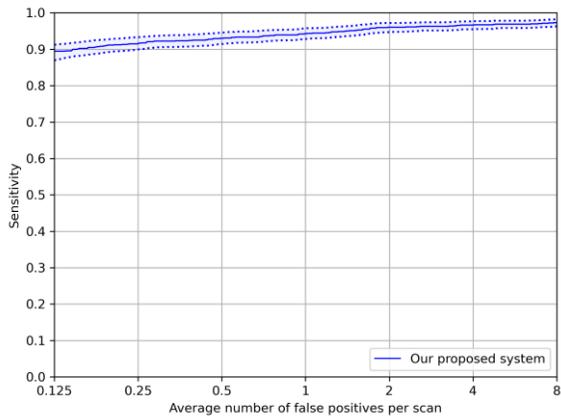

Fig. 4. Free-response receiver operating characteristic (FROC) curves of our proposed system in different configurations. The performance was computed based on 95% confidence interval using bootstrapping with 1,000 bootstraps.

used the Competition Performance Metric (CPM) [43], which calculates the average sensitivity at seven false positive rates (1/8, 1/4, 1/2, 1, 2, 4, and 8 FPs/scan) in the free-response receiver operating characteristic (FROC) curve for assessment [42]. After ten-fold cross-validation, the predictions for all ten test sets were combined to compute the performance and 95% confidence intervals on the full dataset, using bootstrapping with 1,000 bootstraps [44].

The proposed scheme is implemented by applying deep learning library of Keras [45] based on a graphics processing unit(GPU), NVIDIA V100.

## IV. EXPERIMENTAL RESULTS

### A. Nodule candidate localization

The performance of the system at nodule candidate detection stage on each plane, as well as the fused results are presented in Table I. The sensitivity acquired by 1 mm axial slices, 1 mm coronal slices and 1 mm sagittal slices is 91.1%, 82.5%, 81.8%, respectively. After we merged the results from various 1 mm slices, the system achieves a sensitivity of 96.1%. Upon combining the results from the 10 mm axial MIP images, the CAD system detects 98.1% of lung nodules. This proves that every plane provides complementary information for nodule candidate localization, especially the axial plane. Normally, a high sensitivity implies many false positives (FPs) from the CAD system. With 1 mm axial slices, 1 mm coronal slices, 1 mm sagittal slices and 10 mm axial MIP images, our proposed method has 38, 33, 40 and 22 FPs/scan, respectively. The false positive rate is 98 FPs/scan after fusing results from three 1 mm planes, whereas the number of false positives per scan is 108 by fusion of candidates from four streams. The summary of detected lung nodules in size and density type according to the Lung CT screening reporting and the data system [46] at the nodule cadidate detection stage is shown in Table II. The main missed nodules are in the small-size group (3-6 mm), there are 3 ground-glass nodules and 12 solid nodules undetected. However, the detection rate of small nodules is still 97.0%. Regarding nodules larger than 6 mm, only 8 nodules are missed and the detection rate is 98.8%. For ground-glass, sub-solid and solid nodules, the detection rate is 90.6%, 100%, 98.2%, respectively.

### B. False positive candidate exclusion

Our developed system in these configurations is assessed by free-response receiver operating characteristic (FROC) curves, as shown in Fig. 4. The system has a sensitivity of 94.2% with 1.0 FP/scan and 96.0% with 2.0 FPs/scan regardless of nodule size. For detection of nodules smaller than 6 mm, the designed CAD system detects 93.4% (95.0%) of these small nodules at an overall false positive rate of 1.0 (2.0) FP/scan. The CPM score of the CAD scheme with varied configurations at the false positive reduction stage is shown in Table III. Applying 1 mm unsegmented axial slices in the size with eight extra pixels has the best CPM score of 0.940 as same as the score when the patch size with four extra pixels was applied. Compared to that, using 1mm segmented axial slices acquires a lower CPM score (0.933). It shows that the remaining lung boundaries can slightly improve the performance for the removal of false positives. Examples of true positive nodules, false negative ones and false positives after false positive reduction are shown in Fig. 5.

### C. Comparison with published nodule detection systems

To benchmark the performance of our complete CAD program, we list the results from other published papers which were obtained on the same dataset. Sensitivities at different false positive rates in other methods are shown in Table IV.

## V. DISCUSSION

We proposed a novel lung nodule detection system based on multiple planes using convolutional neural networks. The aim of this study was to improve the performance of the deep learning-based CAD system for automatic pulmonary nodule

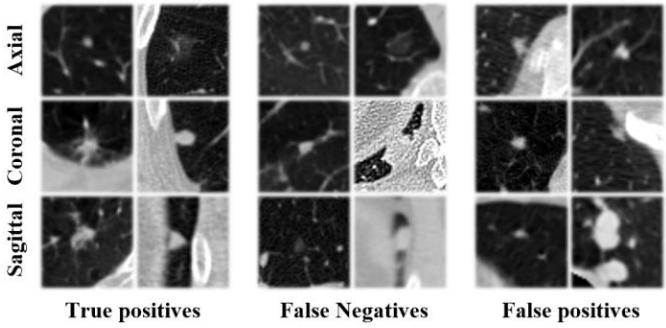
Fig. 5. Examples of true positive nodules, false negative ones and false positives.

detection. Our method achieved comparable performance among the CAD systems evaluated on the LIDC/IDRI database. The combined results from three planes showed better performance than the result from any individual single plane, indicating different planes can provide complementary information for lung nodule detection.

Nodule detection performance was evaluated on the axial, sagittal and coronal planes separately. The axial plane outperforms the rest of two planes in 1 mm slice thickness, achieving a detection sensitivity of 91.1%. In a study regarding human reader performance, it was found that radiologists also have a higher sensitivity, but more false positives for nodule detection on the axial plane compared to the coronal one [47]. In clinical practice, the sagittal plane might be the last option for radiologists to find nodules since vessels tend to be presented as cross sections in this direction. The section of vessels can result in more suspicious findings during reviewing. Through experiments, our study found most of the false positive candidates on the sagittal plane. When we fused the results from three 1 mm planes and 10 mm axial MIP images, the sensitivity increased from the lowest sensitivity of 81.8% to 98.1%, although the number of false positives increased. This suggests that incorporating multiple planes can be an effective approach for 2-D nodule detection. At the false positive reduction stage, we also found that leaving the lung parenchyma unsegmented and using a larger region of interest of extra four pixels in radius boosts the performance of classification. This implies that CNNs can be more accurate to differentiate nodules and false positive finding with more surrounding information.

In a recent study it was shown that detection of small nodules (i.e. nodules with a diameter < 6 mm) is the main challege for which the sensitivity of CAD systems is diffcult to improve [18, 21, 23]. We analyzed detected nodules at the candidate localization stage. Our method had a sensitivity of 97% on detection of these small nodules in various types. There are only 15 out of 502 lung nodules still missed by our method. The detection rate of these small nodules is high since U-net++ can efficiently extract features not only in small receptive fields, but also large receptive fields. Intestingly, there is no missed subsolid nodule. The reason might be that unlike solid nodules, having non-solid parts helps sub-solid ones to be easier differentiated from section of vessels. Moreover, their morphology is more distinguishable compared to ground-glass nodules for convolutional neural networks. Note that the proposed method found 99% of nodules (> 6mm) in large morphological variations. However, there are still some missed nodules. These nodules may appear in unusual locations or close to tissues, which makes detection more difficult for the system.

Recent published approaches on the LUNA16 challenge were summarized in Table IV. To compare the results using the same criteria, we only listed the methods which used the competition performance metric (CPM). Our designed method was ranked third and had the highest sensitivity when the number of false positives allowed is equal to 0.125 FP/scan. The top one is from the work of Setio et al. [21]. With gaining benefits from different CAD systems, they have a better sensitivity when more false positives are allowed. The CPM score from Zhang et al. [23] is also higher and detecting all possible nodule candidates gives them a good upper-bound quality for the false positive exclusion stage. The work by Ozdemir et al. [22], Wang et al. [18], and our previous study [24] demonstrates that a high sensitivity can be achieved, but the large number of false positives per scan that are generated incurs extra reading time for radiologists. The CAD system we propose here shows good performance in detecting these small nodules even after the false positive reduction stage, representing a higher sensitivity than radiologists' [25-27]. We also improve our performance in small nodule detection compared to our previous work [24] (sensitivity: 93.4% vs. 90.4%, at 1.0 FP/scan; sensitivity: 95.0% vs. 91.6%, at 2.0 FPs/scan). Another method from Dou et al. [19] and Xie et al. [20] might need to further improve the discrimination between nodules and wrong findings.

There are some suggestions for the future work. Although this developed CAD system had good performance on this large public dataset, more evaluations on lung cancer screening programs need to be validated. Another interesting topic is that with larger memory in GPUs, convolutional neural networks are capable to be trained by 3-D lung volumes for nodule detection. The system might achieve better performance since vessels and pulmonary nodules can be easily differentiated in 3-D space.

## VI. CONCLUSION

We have developed a multi-planar nodule detection system using convolutional neural networks. The promising performance has shown the effectiveness of combining results from three planes for the candidate detection task. Sharing multi-scale features helped dense convolutional neural networks to become more effective at removing false positives. It would also be essential to assess the generalizability of the proposed system in lung screening trials.